\documentclass[letterpaper, 10 pt, conference]{ieeeconf}  
\usepackage{graphicx}
\usepackage{amsmath,amssymb}
\usepackage{subfigure} 
\usepackage{array}
\usepackage{lipsum}
\usepackage{mathtools}
\usepackage{booktabs}
\usepackage{mwe}
\usepackage{xcolor}
\usepackage{float}
\newtheorem{defi}{Definition}
\newtheorem{remm}{Remark}
\newtheorem{prop}{Proposition}
\newenvironment{proposition}{\begin{prop}\rm }{\hfill \hspace*{1pt} \hfill $\lrcorner$ \end{prop}}
\newenvironment{definition}{\begin{defi}\rm }{\hfill \hspace*{1pt} \hfill $\lrcorner$ \end{defi}}

\IEEEoverridecommandlockouts                              
\title{\LARGE \bf
Qualitative behavior and robustness of dendritic trafficking 
}

\author{Saeed Aljaberi, Timothy O'Leary,  Fulvio Forni
\thanks{S. Aljaberi is supported by Abu Dhabi National Oil Company (ADNOC). T O'Leary is supported by ERC grant StG 716643 FLEXNEURO.
S. Aljaberi, T. O'Leary, and F. Forni are with the Department of Engineering, 
University of Cambridge, CB2 1PZ, UK {\tt sa798|timothy.oleary|f.forni@eng.cam.ac.uk}}
}
\begin{document}

	\maketitle
	\thispagestyle{empty}
	\pagestyle{empty}

	\begin{abstract}
		The paper studies homeostatic ion channel trafficking in neurons. We derive a nonlinear closed-loop model 
		that captures active transport with degradation, channel insertion, average membrane potential activity, and integral control. 
		We study the model via dominance theory and differential dissipativity to show when steady regulation gives way to pathological oscillations. 
		We provide quantitative results on the robustness of the closed loop behavior to static and dynamic uncertainties, which allows us
		to understand how cell growth interacts with ion channel regulation. 
	\end{abstract}
	
	\section{Introduction}

Neurons maintain an array of nonlinear conductances that are mediated by voltage sensitive ion-permeable proteins called ion channels. The lifetime of an ion channel is hours to days, while a neuron typically lives for the lifetime of an animal \cite{marder2006variability}. Ion channels are therefore continually replenished throughout a neuron. This process operates in closed loop: electrical activity is sensed over long timescales and channel densities are controlled by negative feedback to homeostatically maintain neurons at a reference (average) activity level \cite{temporal2014activity,o2010homeostasis}. Homeostasis is essential for maintaining the electrical signaling properties of neurons in the face of biological noise, uncertainty and environmental perturbations, but the underlying control architecture and constraints are poorly understood \cite{o2011neuronal}.

Ion channel homeostasis faces significant constraints due to the complex geometry of neurons \cite{bressloff2009cable,williams2016dendritic}. Channel mRNAs and protein subunits are synthesized at a central location in the cell body but need to be distributed over an extensive dendritic tree. This is achieved by motor proteins via active intracellular transport along a network of filaments called microtubules that span the dendritic tree \cite{kapitein2011way,nirschl2017impact,zarai2017deterministic,bressloff2009cable,williams2016dendritic}. Although this allows newly synthesized mRNAs and proteins to reach the extremities of a neuron, the trafficking mechanism is thought to be largely blind to the identity of the cargo and the location in the neuron. Cargo is trafficked in bulk throughout the network and selected by local sequestration at sites where it is needed. This demand-driven model, known as the {\em sushi belt model} \cite{doyle2011mechanisms}, can be modeled mathematically as a compartmental system \cite{zarai2017deterministic,bressloff2009cable,earnshaw2006biophysical,williams2016dendritic}.

The closed loop behavior of this system critically depends on the dynamics of the transport system and the geometry of the transport network within the neuron. Furthermore, each step in this process is subject to nonlinearities and uncertainties that are inherent in biological systems. It is therefore challenging to make precise statements about the closed loop behavior of ion channel regulation using conventional system-theoretic tools. Neurons undergo physical changes, such as growth, which force them to accommodate changes to a number of physiological properties. Some properties, like settling time, need to stay within a certain range. Other properties, like mRNA synthesis rate, need to increase/decrease depending on the state of the neuron. Indeed, noise and uncertainties are prevalent in biological systems, which requires strong robustness of the feedback regulation mechanisms.

To analyze the closed loop behavior we adopt the novel approaches of dominance theory and differential dissipativity  \cite{Forni2019}, \cite{Miranda-Villatoro2019}.
We study both regimes of steady regulation and pathological oscillations. We illustrate how key parameters affect the behavior, in particular how the size of
the dendritic tree limits the control action, thus the overall regulation performance. We provide a detailed robust analysis of the regulation mechanisms to 
parameter uncertainties and unmodeled dynamics. Our analysis is developed in the nonlinear setting and 
shows that integral control can deal with substantial model uncertainties at the cost of reducing performance, as expected from classical robust control theory
\cite{Desoer1975,Zhou1995,VanDerSchaft1999}.

In Section \ref{model} we derive the model and state the biological assumptions, and show the possible behaviors.  Section \ref{dominance} recalls the main results of dominance theory and p-dissipativity. We provide a classification of possible behaviors in Section \ref{sec:nominal_p}. Section \ref{robust} studies the robustness of the model, to static and dynamic uncertainties. The problem of growth is also discussed. Conclusions follow. 

\section{Compartmental model of dendritic trafficking}
\label{model}

\subsection{Model formulation and biological assumptions}

\begin{figure}[htbp]
	\begin{center}
		\includegraphics[scale=0.7]{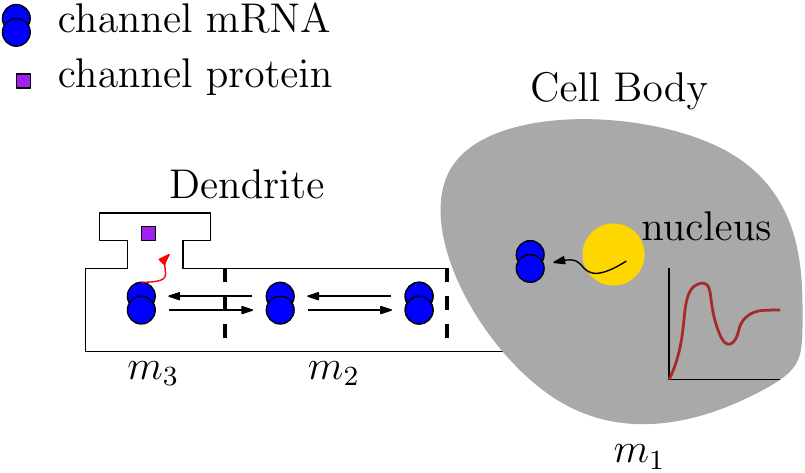}    
		\caption{Role of dendritic trafficking in neural activity.} 
		\label{fig:phen}
	\end{center}
\end{figure}

For expository purposes we consider a neuron as a $3-$compartment system, shown in Figure \ref{fig:phen}. We show later how our analysis is robust to the number of compartments. The first compartment represents the cell body/soma, the second and third compartments represent a dendrite, with the third compartment being the site of cargo usage. Cargo consists of ion channel precursor ($m_i$, for `mRNA') which is produced only in the soma and transformed into functional ion channels ($g_i$) in the final compartment. We note that there is mixed biological evidence for whether the trafficked precursor is mRNA or protein. It is possible that mRNAs are trafficked, then channel protein is synthesized locally. Alternatively, channels are synthesized at the cell body then trafficked. This does not affect our model at our chosen level of abstraction, so we simply refer to the precursor as mRNA.

We model mRNA transport by considering how microscopic active transport along microtubules affects concentration $m_i$ in each compartment. Inspired by \cite{reuveni2011genome}, we derived a mean-field approximation of the random process governed by a \textit{Simple Exclusion Principle} \cite{spohn2012large} taking into account crowding effects of cargo particles. This results in the following nonlinear $3-$compartmental model, with finite capacity compartments, describing dendritic trafficking 
\vspace{-5mm} 			
\begin{align}
\dot{m}_{1} = \
& u(c \!-\! m_{1}) - v_f m_{1}(c \!-\! m_{2}) \nonumber 
+ v_b m_{2}(c \!-\! m_{1})  -\, wm_{1} \nonumber \\
\dot{m}_{2} = \
& v_f m_{1}(c -m_{2})- v_b m_{2}(c  - m_{1}) \nonumber \\
& + \, v_b m_{3}(c  - m_{2})  - v_f m_{i}(c-m_{3}) - wm_{2} \nonumber \\
\dot{m}_{3}  = \
& v_f m_{2}(c -m_{3}) - v_b m_{3}(c -m_{2}) - wm_{3} \label{EQ:m}\\
\tau_{g}\dot{g} = \
&m_{3} - g \label{EQ:V_g}.
\end{align}
In \eqref{EQ:m}, $m_{i} \in [0,c]$ is mRNA concentration in the $i^{th}$ compartment, bounded above by a capacity, $c$; $v_{f}$, $v_{b}$, $w$ are forward velocity, backward velocity, and mRNA degradation, respectively; $u$ is a control input that regulates mRNA synthesis.  \eqref{EQ:V_g} models the dependence of channel protein, $g$, on mRNA concentration in the third compartment: $\tau_g$ is a time constant corresponding to protein synthesis.
		\vspace{-5mm}
\begin{figure}[htbp]
	\begin{center}
			\vspace{2mm}
		\includegraphics[scale=0.4]{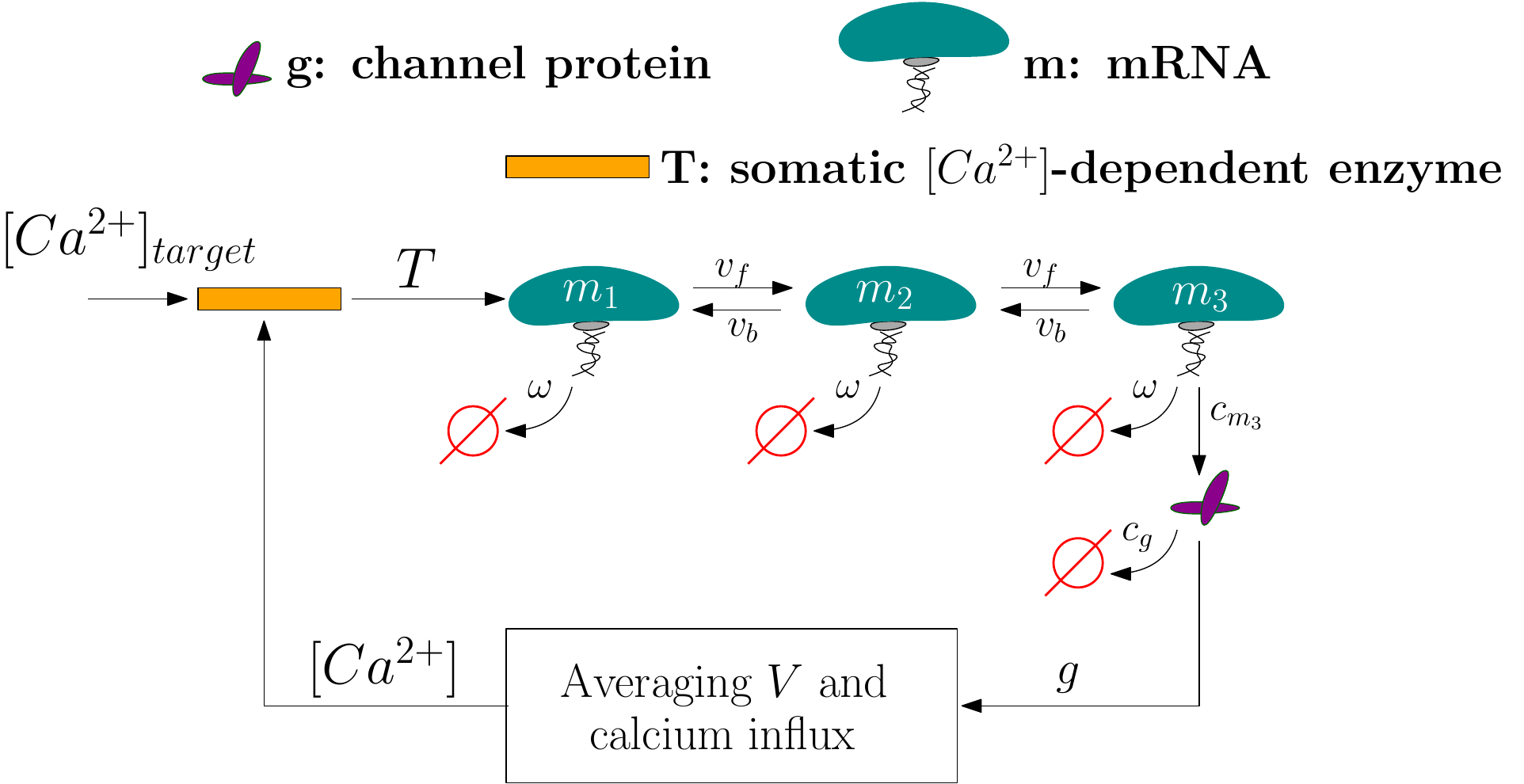}   
		\caption{Closed-loop model of regulation of dendritic trafficking.} 
		\label{fig:homeo}
		\vspace{-5mm}
	\end{center}
\end{figure}
In line with experimental data and existing models \cite{o2014cell,o2013correlations,o2010homeostasis}, we assume that channel mRNA synthesis, which occurs only in the first compartment, is dependent on calcium concentration, $[Ca^{2+}]$, (Figure \ref{fig:homeo}). Existing models posit that biochemical pathways modulate mRNA synthesis according to the deviation of calcium concentration from an effective set-point, $[Ca^{2+}]_{\text{target}}$. The form of the dynamics of the control signal, $T$, that transforms the error signal, $e = [Ca^{2+}]_{\text{target}} - [Ca^{2+}]$ into, $u$, in Equation \eqref{EQ:m} is the subject of ongoing research. In particular, the question of whether perfect set point tracking is achieved is of great interest. Here we consider a biologically plausible controller implementing leaky integral control:
	\begin{equation}
	\label{EQ:leaky-integrator}
	u  = T \ ,\qquad\quad
	\tau_{T}\dot{T} =e - \gamma T - \vartheta(T)
	\end{equation}
where $\gamma > 0$ sets the degradation rate of $T$ and $\vartheta(T)$ enforces positivity and boundedness of $T$. Here, for simplicity, we take 
$\vartheta(T) = a\tan\left(\frac{\pi}{c_{T}}\left(T - \frac{c_{T}}{2}\right)\right)$, for $0 < a \ll 1 $. In the limit of $\gamma \rightarrow 0$,  (\ref{EQ:leaky-integrator}) becomes a pure integrator. 
\begin{figure}[htbp]
		\begin{center}	
	\vspace{-2mm}
	\includegraphics[width=1.2in]{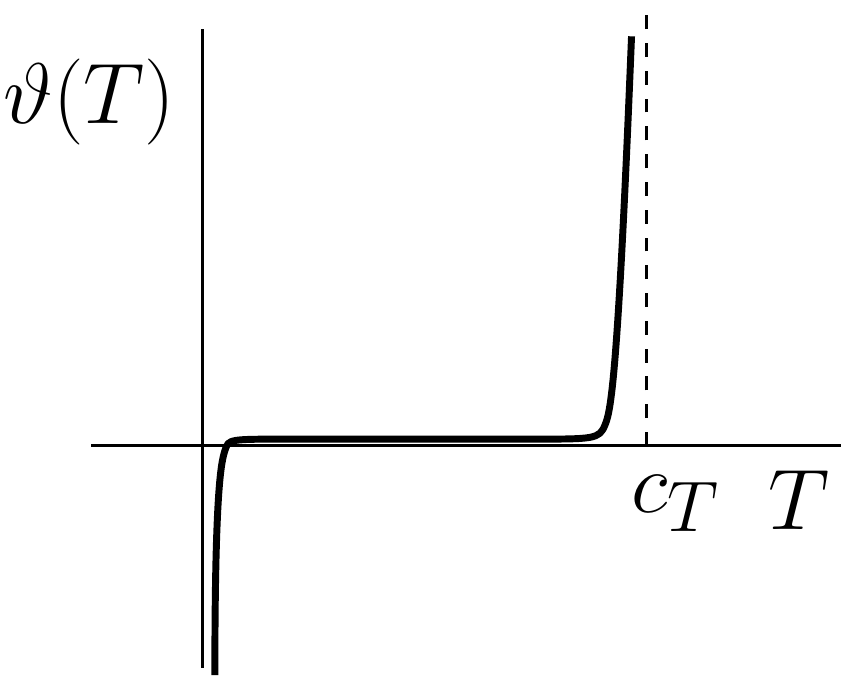}
	\caption{A typical function $\vartheta(T)$}
	\label{fig:sat}
	\vspace{-5mm}
		\end{center}
\end{figure}
	
Calcium concentration varies due to voltage-dependent channels and may be related to the (quasi-steady state) membrane potential via a saturating monotonic relationship: $[Ca^{2+}] =  \frac{\alpha}{ 1 + \exp{V/\beta}}$ with parameters $\alpha$, $\beta$ that capture calcium buffering and the voltage sensitivity of calcium channels \cite{o2013correlations}.

Finally, to model the effect of channel protein concentration on the membrane potential, $V$, we consider the standard single compartment membrane equation, $C\dot{V} {}={}    g_{leak}(E_{leak} - V) + g(E_{g} - V)$, where $C$ is membrane capacitance, $g_{leak}$ is a fixed, leak conductance and the $E_{leak,g}$ terms are equilibrium potentials for each type of ionic conductance. By using a single compartment membrane equation we are assuming that the neuron is equipotential ($V$ is independent of compartment index). We further assume timescale separation between the fast voltage fluctuations and the mRNA synthesis and trafficking mechanisms. We therefore set the membrane potential to its quasi-steady state, $V := V_{\text{ss}} =  \frac{g E_{g} + g_{leak} E_{leak}}{g_{leak} + g}$.
\subsection{Model behavior and nominal parameters}
\begin{figure*}[htbp]
	
	\subfigure[$\tau_{T} = 1000$.]
	{
		\includegraphics[width=2.2in]{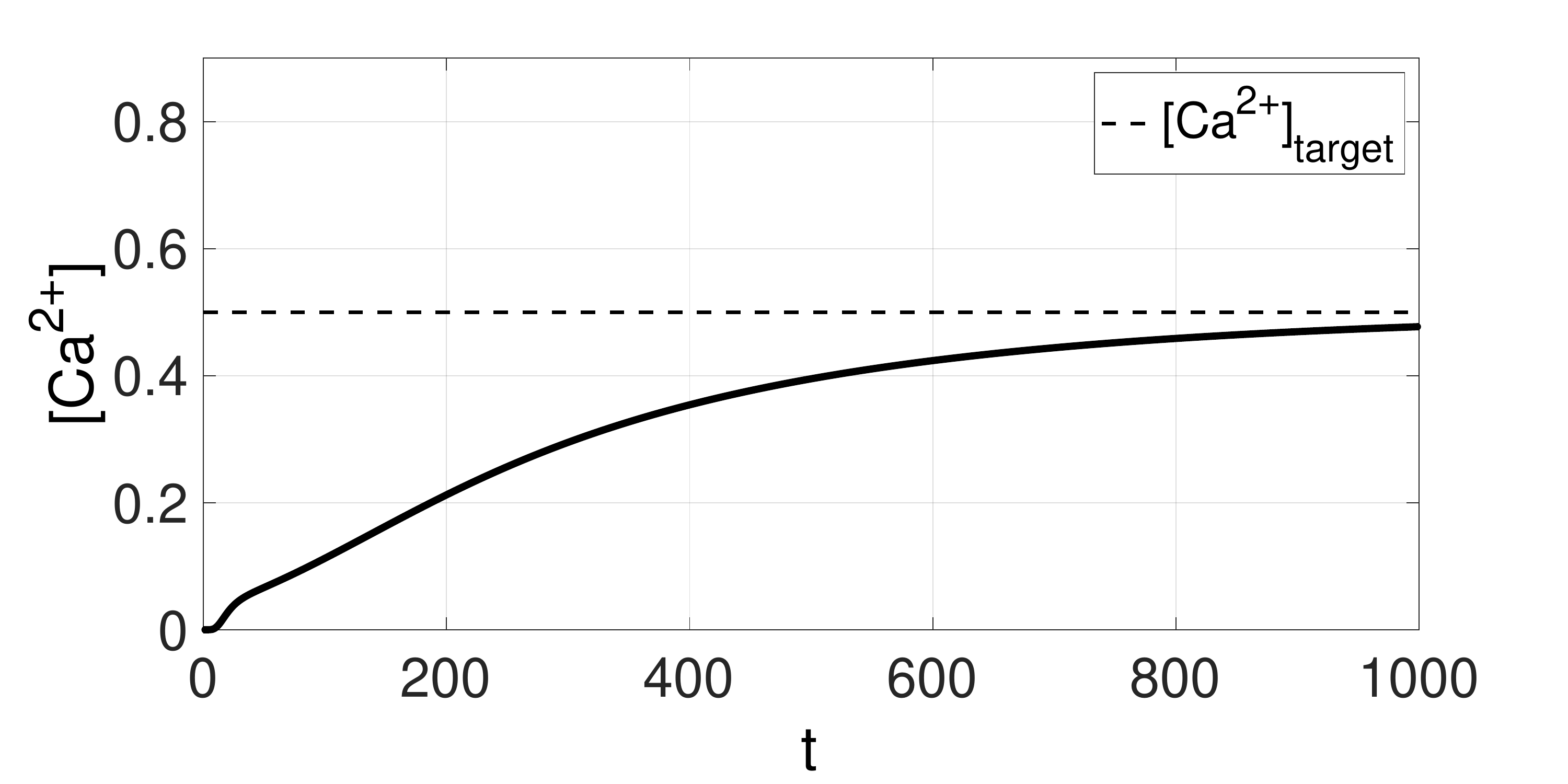}
		\label{fig:LI_1}
	}
	\subfigure[$\tau_{T} = 80$.]
	{
		\includegraphics[width=2.2in]{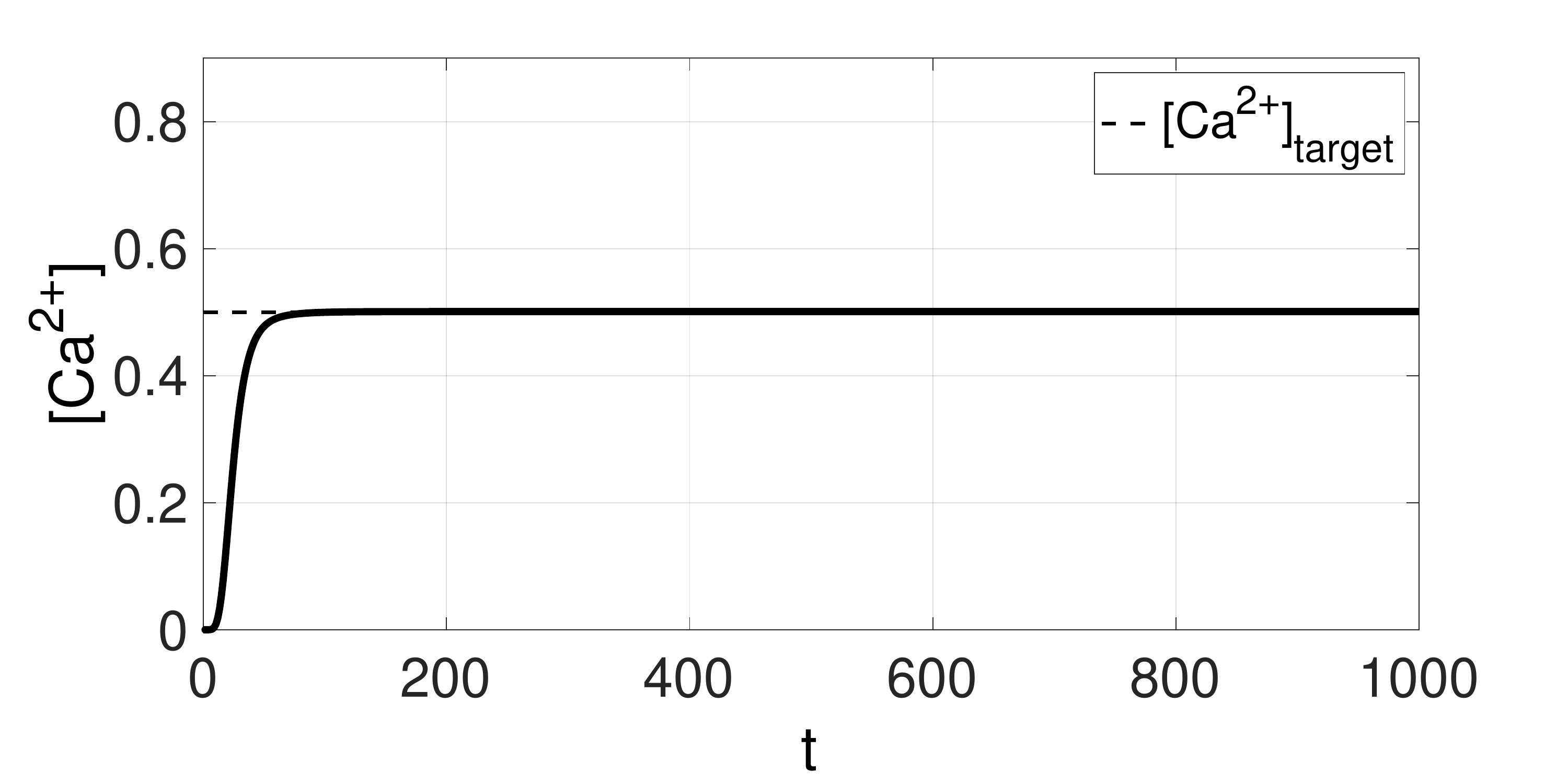}
		\label{fig:LI_2}
	}
	\subfigure[$\tau_{T} = 5$.]
	{
		\includegraphics[width=2.2in]{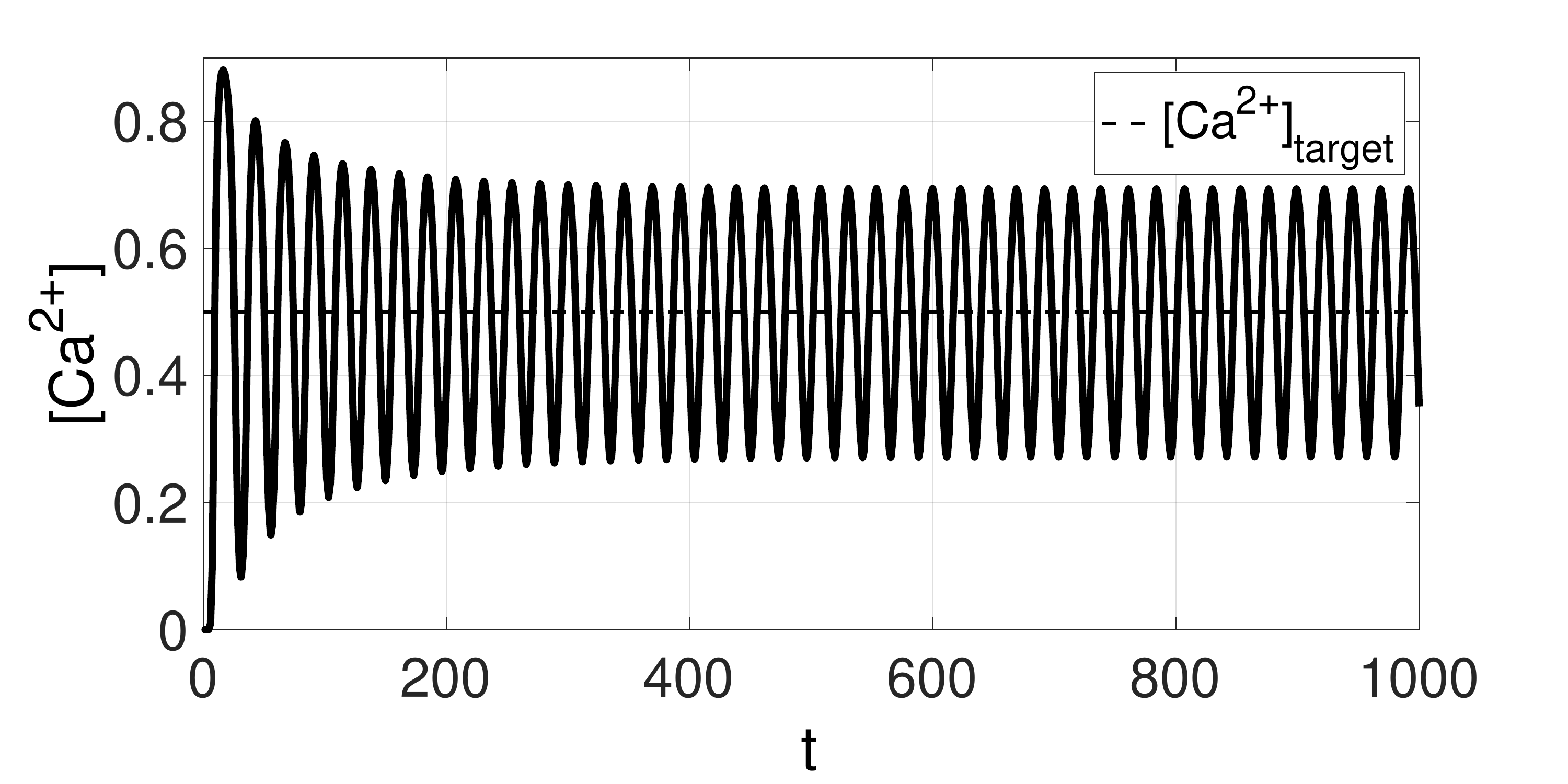}
		\label{fig:LI_3}
	}
	
\subfigure[$\tau_{T} = 1000$. ]
{
	\includegraphics[width=2.2in]{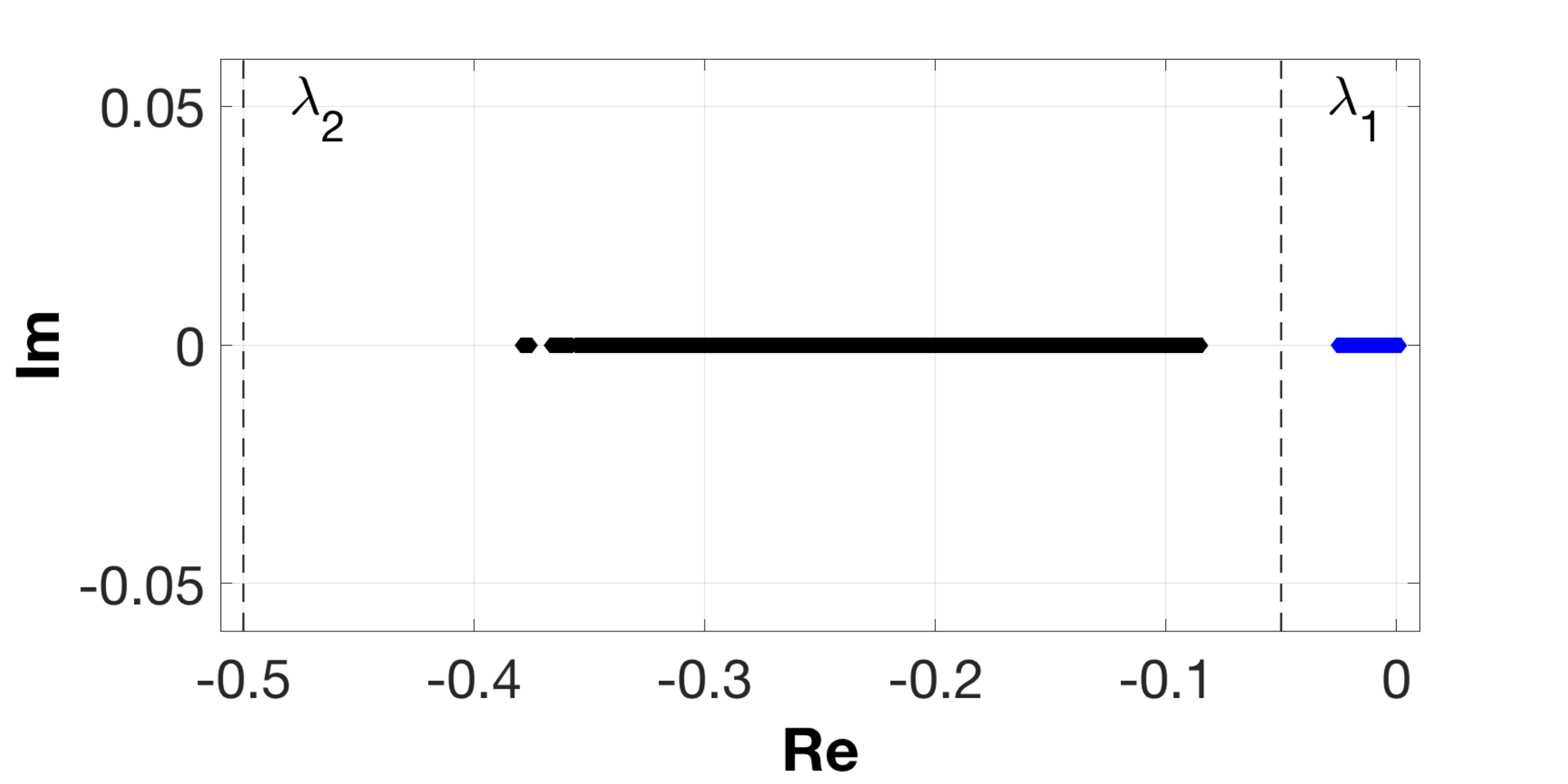}
	\label{fig:roots_LI_1}
}
\subfigure[$\tau_{T} = 80$. ]
{
	\includegraphics[width=2.2in]{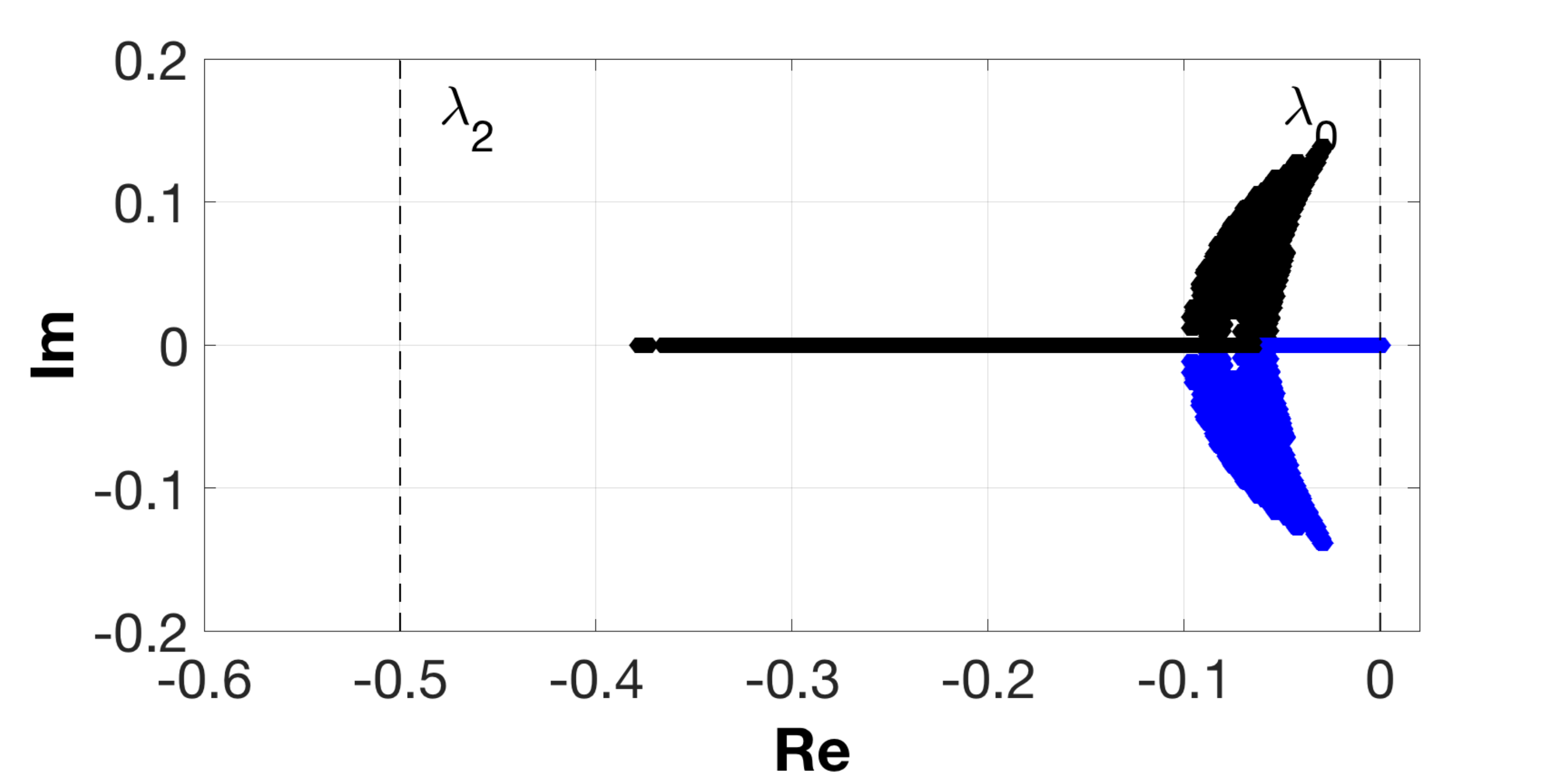}
	\label{fig:roots_LI_2}
}
\subfigure[$\tau_{T} = 5$.]
{
	\includegraphics[width=2.2in]{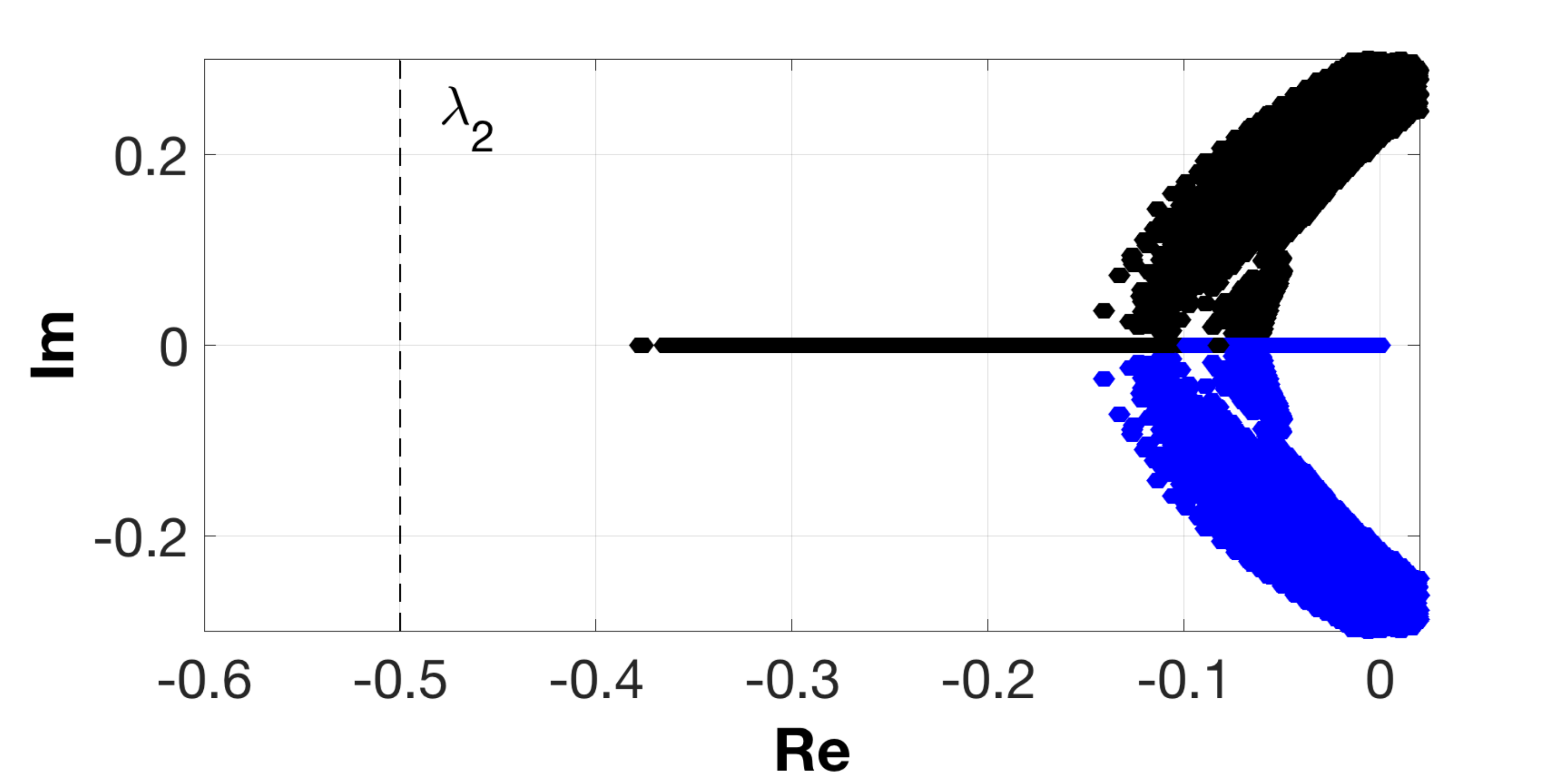}
	\label{fig:roots_LI_3}
}
	\caption{\protect\subref{fig:LI_1}-\protect\subref{fig:LI_3} Response of $[Ca^{2+}]$ for different values of $\tau_{T}$ where the dendritic trafficking model was simulated with the nominal parameter values in Table \ref{table:nominal} and Initial condition $x_{0} = [0.2; 0.2; 0.2; 0.2; 0.2]^{T}$.\protect\subref{fig:roots_LI_1}-\protect\subref{fig:roots_LI_3} two right-most eigenvalues of the Jacobian of the closed loop given by \eqref{EQ:m}-\eqref{EQ:leaky-integrator}. For Figure \protect\subref{fig:roots_LI_2} the right-most eigenvalue does not cross the imaginary axis. Spectra were obtained by sampling $0.1 \leq x_{i} \leq 0.9$, where the black and blue dots depict the movement of the two right-most eigenvalues.}
	\label{fig:P_I}. 
\end{figure*}

\vspace{-4mm}
\begin{table}[htbp]
	\caption{Nominal Parameter Values}
	\vspace{-4mm}
	\label{table:nominal}
	\begin{center}
		\begin{tabular}{ |c|c|c|c| } 
			\hline
			$v_f=1$ & $v_b=0.5$ & $w=1$& $g_{leak}=0.25$ \\ 
			$E_{leak}=-50$ & $E_{g}=20$ & $\alpha=1$& $a=0.0001$  \\
			 $\beta=1$ & $c_{T}=10$ & $\tau_{g}=1$ &  $\tau_{T}=$varies  \\
			 $[Ca^{2+}]_{\text{target}} = 0.5$ & $c=1$ & $\gamma=0$ &  $n=3$  \\
			\hline
		\end{tabular}
	\end{center}
\end{table}
\vspace{-5mm}
Using the nominal parameter values in Table \ref{table:nominal}, 
Figures \ref{fig:LI_1}-\ref{fig:LI_3} summarize the behavior of \eqref{EQ:m},\eqref{EQ:V_g},\eqref{EQ:leaky-integrator} for different values of the integration constant $\tau_T \in \{1000,80,5\}$. Stable regulation is achieved for large integrator time constant $\tau_T$ (slow feedback). Performance improves for smaller time constants (fast feedback). However, performance rapidly degrades with the occurrence 
of pathological oscillations when the integral feedback becomes too aggressive ($\tau_T=5$). A nonzero $\gamma$ in \eqref{EQ:leaky-integrator} will lead to imperfect tracking.  
The analysis in Section \ref{sec:nominal_p} shows that these simulations capture the generic robust behavior of the closed loop system.
	\section{Dominance theory in a nutshell}
	\label{dominance}
	A $p$-dominant linear system with rate $\lambda \geq 0$ has exactly $p$ slow/dominant modes, whose decay
	rate is slower than $-\lambda$, and  $n-p$ fast decaying modes, where $n$ is the system dimension.
	The trajectories of the system rapidly converge to a $p$-dimensional invariant
	subspace capturing the steady-state of the system. In state space representation
	$\dot{x} = A x$, $A\in \mathbb{R}^{n \times n}$, linear $p$-dominance with rate $\lambda$ 
	is certified by the Lyapunov inequality
	$
	A^T P + P A + 2 \lambda I < 0
	$	
	constrained to symmetric matrices $P$ with inertia $(p,0,n-p)$, that is, $p$ negative eigenvalues
	and $n-p$ positive eigenvalues. $p$-dominance can be extended to nonlinear 
	systems of the form $\dot{x} = f(x)$ using the system linearization 
	$\dot \delta x = \partial f(x) \delta x$ along arbitrary trajectories \cite{Forni2019}	
	($\partial f(x)$ is the Jacobian of $f$).
	\begin{definition}	
	A nonlinear system $\dot{x} = f(x)$ is 
	$p$-dominant with rate $\lambda \geq 0$ if there exist a symmetric matrix $P$ with 
	inertia $(p,0,n-p)$ and a positive constant $\varepsilon$ such that 
	\begin{equation}
	\label{eq:LMI_dominance} 
	\partial f(x)^T P + P \partial f(x) + 2\lambda P \leq - \epsilon I 
	\end{equation}
	for all $x \in \mathbb{R}^n$.
	\end{definition}	
	\eqref{eq:LMI_dominance} provides a tractable condition for $p$-dominance
	through convex relaxation, as shown in \cite[Section VI.B]{Forni2019} and \cite[Chapter 4]{lmibook}.
	It enforces a uniform splitting among the eigenvalues of $\partial f(x)$ into $p$ slow eigenvalues
	to the right of $-\lambda$ and $n-p$ fast eigenvalues to the left. 
	Our interest in the property stems from the fact that 
	$p$-dominance strongly constrains the system asymptotic behavior, as clarified by the 
	following proposition from \cite[Corollary 1]{Forni2019}
	\begin{proposition} 
	\label{prop:attractors}
	Every bounded trajectory of a $p$-dominant system $\dot{x} = f(x)$,
	$x \in \mathbb{R}^n$, asymptotically converges to
	
	-~a unique fixed point if $p = 0$;

	-~a fixed point if $p = 1$;

	-~a simple attractor if $p=2$, that is, a fixed point, a set of fixed points and connecting arcs, or a limit cycle. 
	\end{proposition}
	\cite[Theorem 2]{Forni2019} shows that the asymptotic behavior of a $p$-dominant system is 
	captured by a $p$-dimensional dynamics, which thus guarantees  simple attractors for $p\leq 2$.
	We observe that a system can be $p_1$-dominant and $p_2$-dominant, $p_1 \leq p_2$, for different 
	rates $\lambda_1 \leq  \lambda_2$. In using the theory, wee are typically interested in finding the 
	smallest degree of dominance, which corresponds to the simplest asymptotic behavior. 

	Differential dissipativity \cite{Forni2019}, \cite{Miranda-Villatoro2019} extends dominance
	theory to open system. We refer the reader to these publications for details.
	We will use the following notion of $p$-gain.	 
	 \begin{definition}
	 An open system $\dot{x} = f(x) + Bu$, $y = Cx$, with input $u$, output $y$, and state $x$, has
	 $p$-gain $\gamma$ with rate $\lambda \geq 0$ if there exist a symmetric matrix $P$ with 
	inertia $(p,0,n-p)$ and a positive constant $\varepsilon$ such that 
	\begin{equation}
	\label{eq:LMI_gain} 
	\left[\begin{array}{cc}
	\partial f(x)^T P + P \partial f(x) + 2\lambda P + C^T C - \varepsilon I & PB \\
	B^T P & -\gamma^2 I
	\end{array}\right] \leq 0
	\end{equation}
	for all $x \in \mathbb{R}^n$.
	\end{definition}	
	A straightforward specialization of \cite[Theorem 4]{Forni2019}, see also \cite{padoan2019dominance}, provides a 
	differential version of the small gain theorem,
	which allows us to use the $p$-gain of a system to characterize its robustness in 
	presence of model uncertainties, as in 
	classical robust control theory \cite{Desoer1975,Zhou1995,VanDerSchaft1999}.
	\begin{proposition} 
	\label{prop:small_gain}
	For $i \in \{1,2\}$, let $\Sigma_i$ be systems with input $u_i$, output $y_i$, and $p_i$-gain
	$\gamma_i$ with rate $\lambda_i = \lambda \geq 0$. 
	If $\gamma_1 \gamma_2 < 1$ then the closed loop system given by $y_1 = u_2$ and $y_2 = u_1$ is 
	$(p_1+p_2)$-dominant.
	\end{proposition}
	
	Proposition  \ref{prop:small_gain} opens the way to the study of robust attractors that are not fixed points.
	This is particularly relevant in system biology. 
	In what follows we will take advantage of the tractability of \eqref{eq:LMI_dominance} 
	combined to Proposition \ref{prop:attractors} to characterize the steady state behavior of 
	dendritic traffic regulation. Then,
	we will use the notion of $p$-gain in combination with Proposition \ref{prop:small_gain} to 
	study its robustness.
\begin{table*}[htbp]
\begin{center}
	\caption{Solutions to \eqref{eq:LMI_dominance} for uncertainties in Table \ref{table:2-dom}.}
	\label{table:p}
\begin{tabular}{ c c }
	\toprule
	\textbf{$1-$dominance} & \textbf{$2-$dominance}\\
	\midrule\\
	\addlinespace[-2ex]
	$P_{1} = \begin{bmatrix}
	0.1372 &   0.0602  &  0.0731  & -0.0043 &  -0.2885 \\
	0.0602  &  0.1256  &  0.0867  & -0.0146 & -0.2571 \\
	0.0731 &   0.0867  &  0.1721  & -0.0644  & -0.2850 \\
	-0.0043 &  -0.0146 &  -0.0644 &   0.0713 &   0.0489 \\
	-0.2885 &  -0.2571 &  -0.2850&    0.0489 & -10.4987
	\end{bmatrix}$ &
$P_{2}	=	\begin{bmatrix}
15.0416 & -45.5028 & -86.9065 &  -4.5915&  -12.0336 \\
-45.5028  &-26.7821 & -81.5167  &-19.3020  & -2.0001 \\
-86.9065 & -81.5167 & -15.7094 & -28.5125 &  28.5969 \\
-4.5915 & -19.3020 & -28.5125  & 28.3913  & 37.2012 \\
-12.0336 &  -2.0001 &  28.5969 &  37.2012 & -43.7541
\end{bmatrix}$ \\
	\addlinespace[1.5ex]
	\bottomrule
\end{tabular}
\end{center}
\end{table*}
\section{Nominal behavior and differential analysis of dendritic trafficking}	
\label{sec:nominal_p}
	 Denoting by $\dot{x} = f(x)$ the closed loop dynamics \eqref{EQ:m}-\eqref{EQ:leaky-integrator}, Figures 	\ref{fig:roots_LI_1}-\ref{fig:roots_LI_3} show the position of the eigenvalues of the Jacobian $\partial f(x)$ 
	for different levels of control aggressiveness, through the selection of values $\tau_{T} \in \{1000,80,5\}$ and $x$, for nominal parameter values in Table \ref{table:nominal}. 
	For readability, we show only the two right-most eigenvalues of the Jacobian. The others are always to the left of $-0.5$.
	Figures \ref{fig:roots_LI_1}-\ref{fig:roots_LI_3} can be roughly separated in two groups: \emph{stable linearization} - $\partial f(x)$ has stable eigenvalues; \emph{Hopf-bifurcation} - $\partial f(x)$ has unstable complex eigenvalues. These two groups explain the difference between stable regulation at steady state and the appearance of oscillations for small $\tau_{T}$. 
	For instance, for large $\tau_T = 1000$ (slow feedback) there are two real, stable eigenvalues, as shown in Figure \ref{fig:roots_LI_1}. This is compatible with the behavior in Figure \ref{fig:LI_1}.  
	As the integrator dynamics becomes faster, $\tau_T = 80$, the two right most eigenvalues coalesce and bifurcate, Figure \ref{fig:roots_LI_2}. Convergence becomes faster, as shown in Figure \ref{fig:LI_2} but
	damped oscillations may appear. Finally, for aggressive feedback, $\tau_T = 5$, the complex unstable eigenvalues 
	in Figure \ref{fig:roots_LI_3} justify the occurrence of sustained oscillations in 
	Figure \ref{fig:LI_3}. 
	
	The connection between Jacobian eigenvalues and closed loop behavior can be made rigorous 
	through dominance analysis, by solving the linear matrix inequality \eqref{eq:LMI_dominance} 
	for $\{m_i, g\} \in [0.1c , 0.9c]$, and using rates $\lambda_0, \lambda_1, \lambda_2$ 
	in Figures \ref{fig:roots_LI_1}-\ref{fig:roots_LI_3}; CVX \cite{grant2014cvx} was used to numerically solve \eqref{eq:LMI_dominance} for each case of $\tau_{T}$.
	The first observation is that the system is always $2-$dominant with rate $\lambda_2 = 0.5$. A common
	solution $P$ can be found for $\tau_T \in [3,3000]$. This has a striking conclusion: the steady state 
	of the closed loop system is compatible with planar dynamics, captured by a simple attractor. 
	This means that for $\tau_T \in [3,3000]$ the closed loop system \emph{either converges to a fixed point or enters into 
	sustained oscillations}. This conclusion can be refined:	
	\begin{itemize}
	\item ($\tau_T = 1000$)
	For slow feedback, $\lambda_1=0.05$ separates the two real eigenvalues into two subgroups as
	shown in Figure \ref{fig:roots_LI_1}. Feasibility of \eqref{eq:LMI_dominance}  shows that 
	the system is $1$-dominant with rate $\lambda_{1}=0.05$, which guarantees \emph{convergence to a fixed point}.
	\item  ($\tau_T = 80$)
	As the integrator dynamics become faster
	the two right most eigenvalues bifurcate and $1-$dominance is lost (Figure \ref{fig:roots_LI_2}).  		However, the system is still $0-$dominant locally with rate $\lambda_{0}=0$. This guarantees \emph{local 
	convergence to the fixed point}. 	
	\item  ($\tau_T = 5$) For aggressive integrator dynamics the system 
	is $2-$dominant with rate $\lambda_{2}=0.5$. It cannot be $1$-dominant (complex right-most eigenvalues)
	and it cannot be $0$-dominant (unstable eigenvalues). $2$-dominance combined with the instability of the 
	fixed point guarantees that \emph{sustained oscillations are the only possible steady state behavior}. 
	\end{itemize}	
		\vspace{-3mm}
	\section{Robustness and growth}
	\label{robust}		
	\subsection{Parametric uncertainties}
	The analysis above shows how the feedback time constant $\tau_T$ affects 
	regulation. We now study robustness to other physiologically relevant parameters (such as velocities and length) using dominance theory,
	looking at the three regimes $\tau_T \in \{1000,80,5\}$. For $\tau_T = 1000$ a stable closed
	loop behavior is preserved for any uncertainty in Table \ref{table:2-dom} (left column). For $\tau_T=5$, the robustness of the oscillatory regime is guaranteed for  parameter ranges specified in Table \ref{table:2-dom} (right column). 
	A local robust analysis is also developed for $\tau_T = 80$. This is a fragile case for dominance analysis, 
	which we address numerically by looking at specific local regions.
			\begin{table}[htbp]
			\caption{}
			
			\label{table:2-dom}
			\setlength{\tabcolsep}{5pt} 
			\begin{tabular}{ccc}
				&  $1-$dom:  $\tau_{T}  \!=\!  1000$, $\lambda \!=\! 0.05$  &   $2-$dom: $\tau_{T}  \!=\!  5$, $\lambda  \!=\!  0.5$ \\\hline
				$v_{f}$ & $[0.2,1.2]$ & $[0.8,1.3]$\\
				$v_{b}$ & $[0.35, 1.1]$& $[0.45, 1.4]$\\ 
				$\tau_{g}$ & $[0.1, 1.4]$ & $[0.8, 1.2]$ \\ \hline
			\end{tabular}
		\vspace{-3mm}
	\end{table}	
	
	For $\tau_{T} = 1000$, the controller guarantees robust $1$-dominance with rate $\lambda = 0.05$
	to uncertainties in Table  \ref{table:2-dom} (left column). Indeed, the matrix $P_1$ in Table \ref{table:p} is a solution to  \eqref{eq:LMI_dominance} 
	for all parametric uncertainties in Table \ref{table:2-dom} (left column).
	Robust stable regulation is thus guaranteed for these uncertainties.
	Stable regulation is also preserved when the velocity constants are replaced by nonlinear functions $v_{f}(m_{i-1}, m_{i}, m_{i+1})$ and  $v_{b}(m_{i-1}, m_{i}, m_{i+1})$ 
	whose slopes $v_f'$ and $v_b'$ belong to the intervals defined in Table \ref{table:2-dom}. 

	For $\tau_T = 5$, the closed loop system is robustly $2$-dominant with rate $\lambda =0.5$ to uncertainties in 
	Table \ref{table:2-dom} (right column). This is certified by the matrix $P_2$ in Table \ref{table:p} which is a solution 
	to \eqref{eq:LMI_dominance} for all parametric uncertainties in Table \ref{table:2-dom} (right column).
	As discussed in other sections,  $2-$dominance is not sufficient to claim robust oscillations. However, the unique equilibrium of the system is always unstable 
	for parameters in Table \ref{table:2-dom} (right column) which, combined with $2$-dominance, guarantees robust oscillations.
	
	For $\tau_{T} = 80$, the closed loop is moving from a stable to an oscillatory regime (complex stable poles in the Jacobian). High sensitivity to parameter variations is thus expected. 
	Table \ref{table:0-d_LI} shows the trade-off between parameter ranges and size of the region of $0-$dominance. 
	\vspace{-2mm}
	\begin{table}[htbp]
		\caption{$0-$dominance: $\tau_{T} = 80$, $\lambda = 0$.\vspace{-2mm}}
		\begin{center}
			\label{table:0-d_LI}
			\begin{tabular}{cccc}
				&  $25\%$ around $x^{*}$ & $20\%$ around $x^{*}$ & $15\%$ around $x^{*}$ \\\hline
				$v_{f}$ & $[0.8,1.2]$ & $[0.55,1.45]$ & $[0.4,1.65]$\\
				$v_{b}$ & $[0.35, 1.1]$ & $[0.2, 1.3]$ &$[0.05,1.5]$ \\ 
				$\tau_{g}$ & $[0.8, 1.2]$ & $[0.55, 1.45]$ & $[0.4,1.6]$\\ \hline
			\end{tabular}
		\end{center}
		\vspace{-5mm}
	\end{table}
	\subsection{Growth}
	\label{sec:growth}
	
	How does a neuron tune its transcription rate in the presence of growth? A bigger neuron requires more biomolecules to be synthesized and their traveling distance is longer. With these variations, can a neuron withstand and maintain a stable nominal behavior? Growth can be modeled in two ways: by increasing the number of compartments or by adapting capacity and velocity parameters. We adopt the latter for simplicity. 
	
\begin{figure}[htbp]
		\begin{center}
			\vspace*{2mm}
			\includegraphics[width=3.0in]{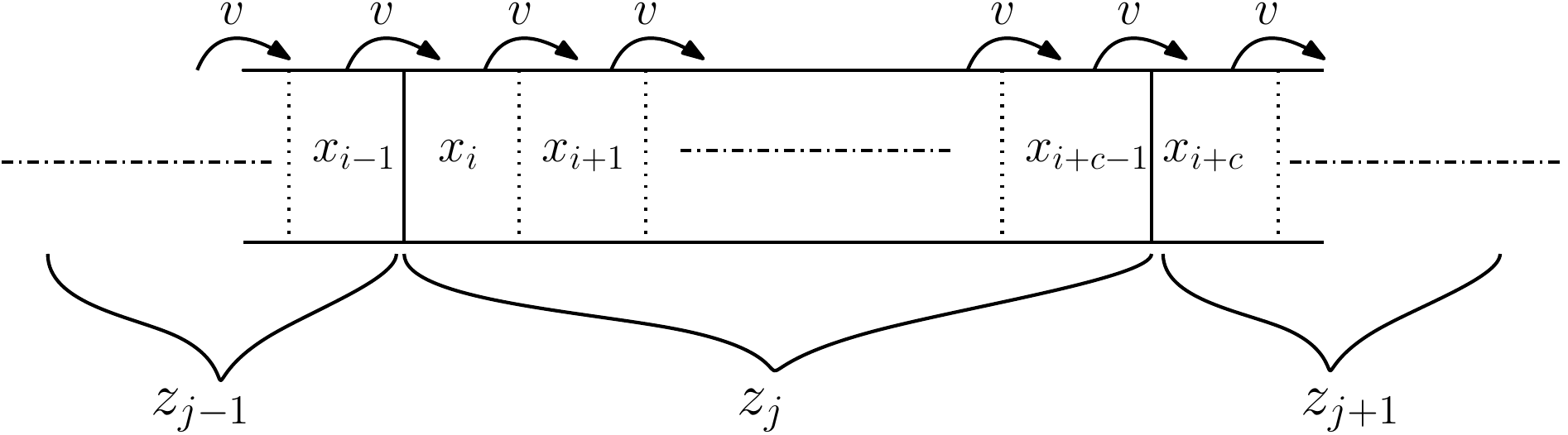}
			\caption{A schematic representation of equations (\ref{EQ:SEP})-(\ref{EQ:SEP_2}).}
			\label{fig:SEP_ext}
		\end{center}
			\vspace{-5mm}
	\end{figure}

\begin{figure}[htbp]
		\begin{center}
			
			\subfigure[$1-$dominance with $\lambda_{1} = 0.05$.]
			{
				\includegraphics[width=1.5in]{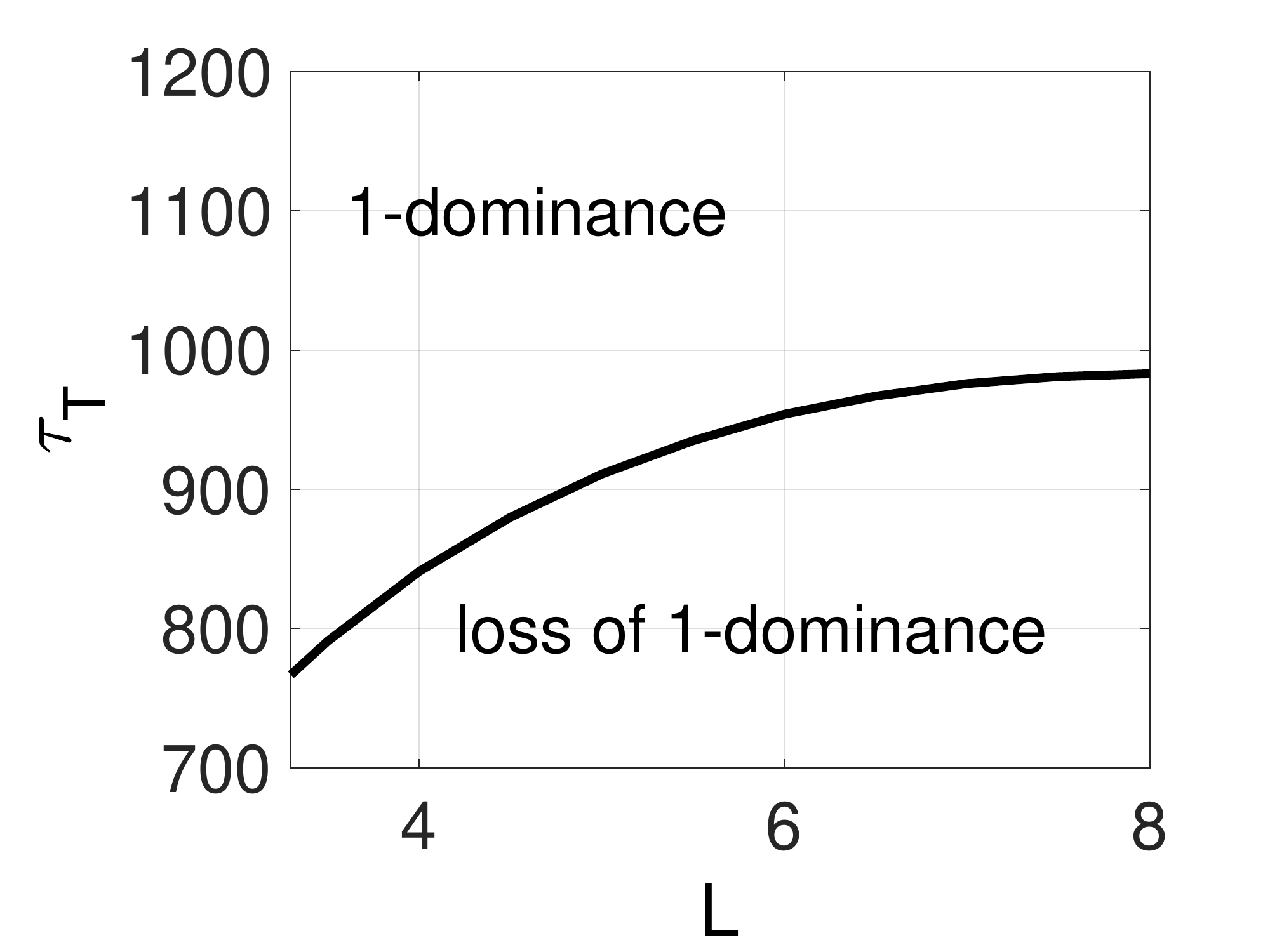}
				\label{fig:1_d_r}
			}
			\subfigure[$2-$dominance with $\lambda_{2} = 0.4$.]
			{
				\includegraphics[width=1.5in]{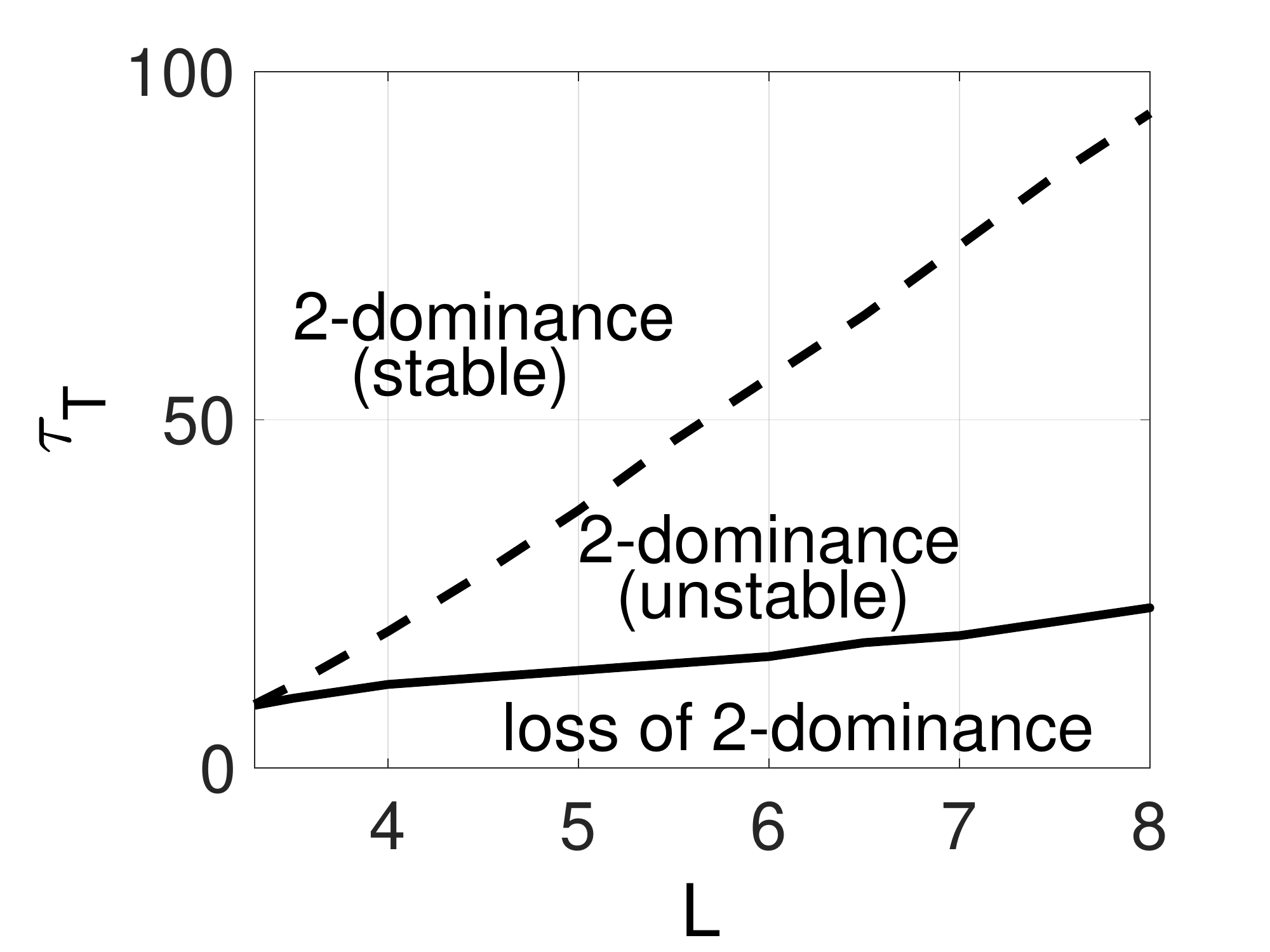}
				\label{fig:2_d_r}
			}
			\caption{Trade-off between $\tau_{T}$ and $L$ for $1-$ and $2-$dominance.}
			\label{fig:growth}
			\vspace{-7mm}
		\end{center}
	\end{figure}

	We consider $1$-dimensional growth, where $L$ represents the neuron's total length. The identity $c= L/n$ relates length $L$ to compartment's capacity $c$ 
	and to compartments number $n$. Growth corresponds to larger $L$ thus larger capacities. Forward and backward speeds are also updated accordingly. Starting from the microscopic picture, suppose that each compartment can fit $c$ number of molecules as shown in Figure \ref{fig:SEP_ext}. The figure shows a large compartment $z_{j}$ of capacitance $c$ and its constituent unit compartments $x_{i}$'s, each of capacity $1$. The rate of change of molecules in compartment $z_{j}$ is given by
	\begin{equation}
	\dot z_{j} = vx_{i-1}(1 - x_{i}) - vx_{i+c-1}(1 - x_{i+c})
	\label{EQ:SEP}
	\end{equation}        
	where the internal exchange of molecules sum to zero. We focus on particles that enter and leave $z_{j}$, assuming that particles are homogeneously distributed and spatially indistinguishable (well-mixed)
	in each compartment $z_{j}$, that is,  		
	\begin{equation}
	x_{i} = x_{i+1} \dots = x_{i+c-1} = \frac{z_{j}}{c}.  
	\label{EQ:SEP_2}
	\end{equation}        
	Substituting (\ref{EQ:SEP_2}) into (\ref{EQ:SEP}), we get
	\begin{align}
	\dot z_{j} = \
	& v\frac{z_{j-1}}{c} \left(1 - \frac{z_{j}}{c} \right) - v\frac{z_{j}}{c} \left(1 - \frac{z_{j+1}}{c} \right) \nonumber \\
	= \
	& \frac{v}{c^2}z_{j-1}\left(c - z_{j}\right) - \frac{v}{c^2}z_{j}\left(c - z_{j+1}\right).
	\label{EQ:SEP_3}
	\end{align}        
	Equation (\ref{EQ:SEP_3}) shows that increasing $L$ the compartment size increases linearly and the velocities scale with $1/c^2$ or equivalently $1/L^2$.
	In summary, growth is modeled by the following parameter scaling in (\ref{EQ:m}): 
	\begin{equation}
	v_{f}   \rightarrow \frac{v_{f}}{c^2} \,,\
	v_{b}   \rightarrow \frac{v_{b}}{c^2} \,,\
	c     = \frac{L}{n} \ .
	\label{EQ:SEP_4}
	\end{equation}
	Within this modeling framework, the question of growth reduces to a question of robustness to parameter variations. The first question is: given an integrator time constant $\tau_{T}$, how much can the neuron grow before loosing stability?  We answer through $1-$dominance, by deriving intervals of length $L$ that guarantee $1$-dominance for a fixed time constant $\tau_{T}$, as shown in  Figure \ref{fig:1_d_r}. As expected, stable regulation for longer neurons requires less aggressive feedback (larger $\tau_T$).
	For any time constant $\tau_T$, there is a threshold length after which $1$-dominance is lost. 
	This regime is characterized by the emergence of damped oscillations, which eventually 
	degrade into sustained oscillations for longer lengths. In fact, Figure  \ref{fig:2_d_r} 
	shows that $2$-dominance of the closed loop is preserved for large variations (both on $L$ or $\tau_T$)
	with limit cycles appearing when the time constant is sufficiently small or the length is sufficiently large, that is, when the equilibrium of the system loses stability. 
	\vspace{-4mm}
	\subsection{Unmodeled dynamics}
	\label{unm_rob}	
\begin{figure}[htbp]
		\begin{center}
			\vspace*{2mm}
			\includegraphics[width=3in]{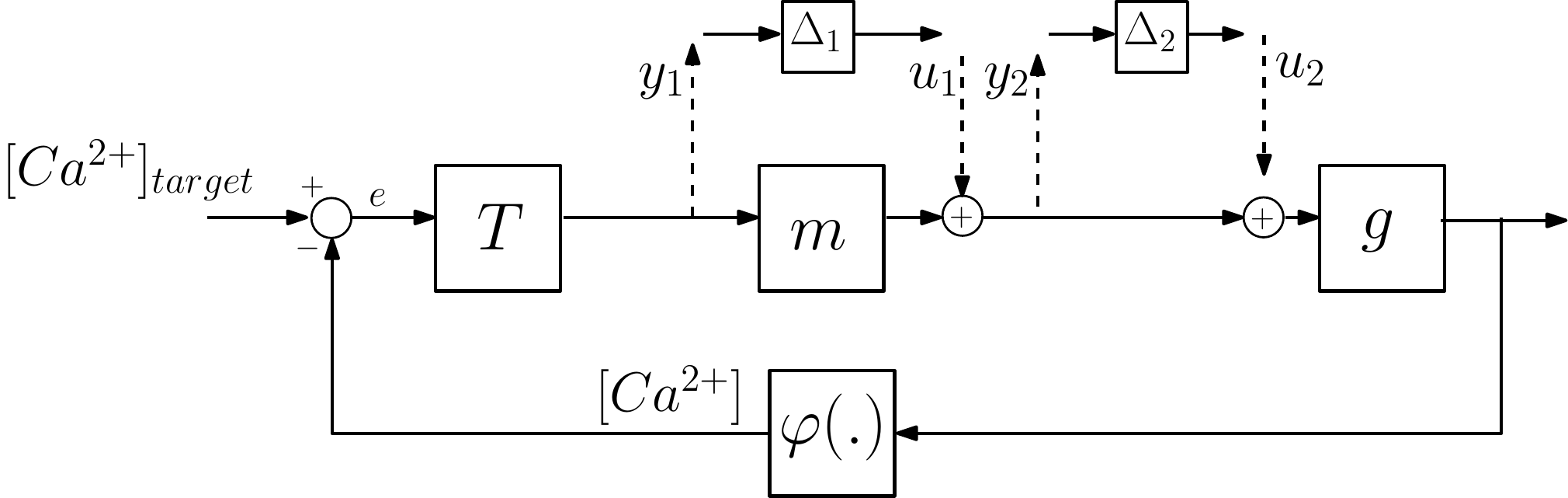} \vspace{-2mm}
			\caption{A schematic showing how the unmodeled dynamics affect the nominal closed-loop as dynamic perturbations.}
			\label{fig:unm_dyn}
		\end{center}
	\end{figure}

	Both growth and parametric variations have been modeled in previous sections as \emph{static} uncertainties. We now consider \emph{dynamic} uncertainties typically arising from unmodeled dynamics and modelling simplifications.We will model these uncertainties as (possibly nonlinear) $0$-dominant dynamic perturbations, $\Delta_1$ and $\Delta_2$, acting on the nominal closed loop as shown in Figure \ref{fig:unm_dyn}. $\Delta_1$ corresponds to additive perturbations, such as neglected transport phenomena. $\Delta_2$  corresponds to multiplicative
	uncertainties such as neglected fast dynamics in protein synthesis. 
	
	We assess the robustness of the closed loop using
	the notion of $p$-gain in Section \ref{dominance} and the small gain interconnection 
	in Proposition \ref{prop:small_gain}, which guarantees that perturbations do not affect the 
	the dominance of the closed loop if the product of the nominal gain and of the perturbation gain is less than one. Indeed, for the nominal parameters in Table \ref{table:nominal}, solving \eqref{eq:LMI_gain}, the nominal closed loop in Figure \ref{fig:unm_dyn} has $1$-gain $\gamma_{cl_{1}} = 0.4549$ from $u_1$ to $y_1$ and and $1$-gain $\gamma_{cl_{2}} = 2.7415$ from $u_2$ to $y_2$, both with rate $\lambda_{1} = 0.05$.
	$1$-dominance of the closed loop, i.e. steady regulation, is thus preserved for any perturbation 
	$\Delta_1$ whose $0$-gain $\gamma_1$  satisfies $\gamma_1 <  1 / {\gamma_{cl_1}}$ with rate $\lambda_{1} $.
	$1$-dominance is also preserved when $\Delta_2$ has $0$-gain $\gamma_2 < {1} / {\gamma_{cl_2}} $ with rate $\lambda_{1}$.
	
\begin{figure}
		\begin{center}
							
			\includegraphics[width=1.8in]{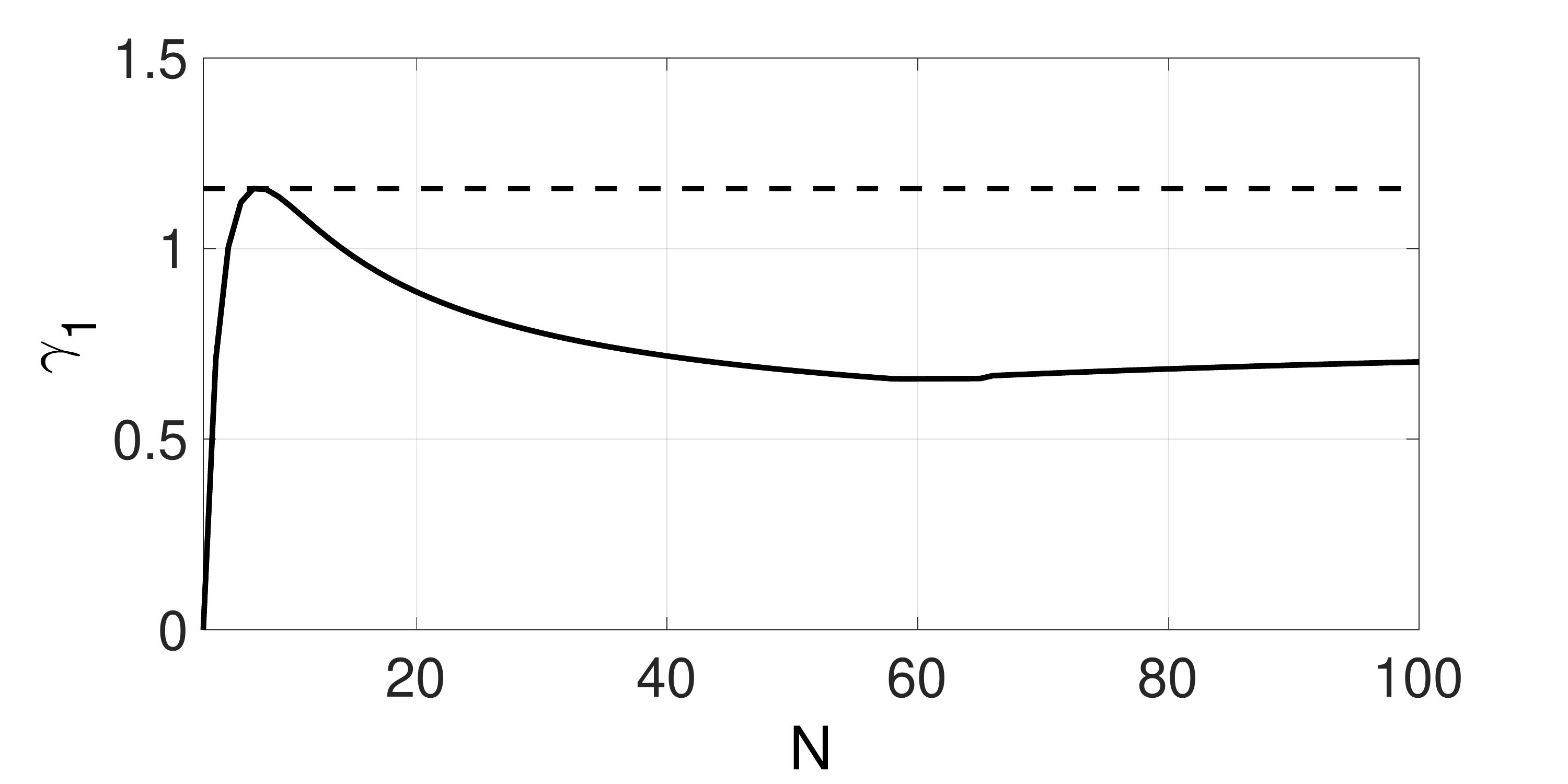}
			\caption{$0$-gain $\gamma_1$ of $\Delta_1 = \Sigma_N - \Sigma_3$ for $3 \leq N \leq 100$ and rate $0.05$.}
			\label{fig:MOR_linear}
			\vspace{-7mm}
		\end{center}
	\end{figure}

	As an example we study closed loop regulation when  the $3$-compartmental model of transport $\Sigma_3$ is
	replaced by a more detailed  $N$-compartmental model $\Sigma_N$, $N>3$. 
	For this case $\Delta_1$ represents the mismatch dynamics $\Sigma_N-\Sigma_3$.
	For simplicity we restrict our analysis to linear transport models, that is, 
	we take $\Sigma_3$ as in (\ref{EQ:m}) but ignore compartment saturation.
	$\Sigma_N$ is also a linear compartmental system. 
	Its parameters are scaled according to Section \ref{sec:growth}, but this time keeping a constant length $L$ and varying the number
	of compartments. Figure \ref{fig:MOR_linear} shows how the $0$-gain $\gamma_1$ (rate $0.05$) of $\Delta_1$
	changes with the number of compartments. For the nominal parameters in Table \ref{table:nominal}, 
	$\gamma_1$ peaks at $1.1575$, which guarantees that the closed loop
	behavior remains unchanged if we replace our $3$-compartmental transport model with
	a more detailed transport model based on $3 \leq N \leq 100$ compartments.	

	A similar analysis can be developed to account for unmodeled protein dynamics to show that 
	sufficiently fast reactions can be safely neglected.
   These examples show the flexibility of the framework in systems biology for capturing
	heterogeneous families of perturbations, mimicking classical robust control. We note
	that the approach is not limited to linear perturbations and can be extended 
	beyond fixed point analysis to study robust oscillations via $2$-dominance. 
\section{Conclusion}
\label{concl}	
We presented a nonlinear model of dendritic trafficking that captures spatial and crowding effects. We studied  integral regulation of dendritic trafficking in the context of ion channel regulation,
providing results on its robustness to parametric and dynamic uncertainties, and to growth. A large number of questions remain unanswered. 
Our analysis of growth, for example, is based on the variation of the dendrite length but neurons develop by branching, expanding their dendritic trees within complex 
morphologies and with varying diameters among different sections. This is an important research direction that we believe can also be addressed with the tools adopted in this paper. 	
Another promising direction is the study of regulation of synapses, or non-homogeneous compartments with different conductance levels. 
Finally, in this study we assumed that the soma is responsible for the neuron's entire regulation process. However, experimental studies suggest that
regulation is achieved by coordination between global (integral control) and local (degradation) mechanisms. This is an intriguing direction that we will
explore in future publications.
	\bibliographystyle{IEEEtran}

\begin{thebibliography}{10}
	\providecommand{\url}[1]{#1}
	\csname url@samestyle\endcsname
	\providecommand{\newblock}{\relax}
	\providecommand{\bibinfo}[2]{#2}
	\providecommand{\BIBentrySTDinterwordspacing}{\spaceskip=0pt\relax}
	\providecommand{\BIBentryALTinterwordstretchfactor}{4}
	\providecommand{\BIBentryALTinterwordspacing}{\spaceskip=\fontdimen2\font plus
		\BIBentryALTinterwordstretchfactor\fontdimen3\font minus
		\fontdimen4\font\relax}
	\providecommand{\BIBforeignlanguage}[2]{{%
			\expandafter\ifx\csname l@#1\endcsname\relax
			\typeout{** WARNING: IEEEtran.bst: No hyphenation pattern has been}%
			\typeout{** loaded for the language `#1'. Using the pattern for}%
			\typeout{** the default language instead.}%
			\else
			\language=\csname l@#1\endcsname
			\fi
			#2}}
	\providecommand{\BIBdecl}{\relax}
	\BIBdecl
	
	\bibitem{marder2006variability}
	E.~Marder and J.-M. Goaillard, ``Variability, compensation and homeostasis in
	neuron and network function,'' \emph{Nature Reviews Neuroscience}, vol.~7,
	no.~7, p. 563, 2006.
	
	\bibitem{temporal2014activity}
	S.~Temporal, K.~M. Lett, and D.~J. Schulz, ``Activity-dependent feedback
	regulates correlated ion channel mrna levels in single identified motor
	neurons,'' \emph{Current biology}, vol.~24, no.~16, pp. 1899--1904, 2014.
	
	\bibitem{o2010homeostasis}
	T.~O'Leary, M.~C. van Rossum, and D.~J. Wyllie, ``Homeostasis of intrinsic
	excitability in hippocampal neurones: dynamics and mechanism of the response
	to chronic depolarization,'' \emph{The Journal of physiology}, vol. 588,
	no.~1, pp. 157--170, 2010.
	
	\bibitem{o2011neuronal}
	T.~O'Leary and D.~J. Wyllie, ``Neuronal homeostasis: time for a change?''
	\emph{The Journal of physiology}, vol. 589, no.~20, pp. 4811--4826, 2011.
	
	\bibitem{bressloff2009cable}
	P.~C. Bressloff, ``Cable theory of protein receptor trafficking in a dendritic
	tree,'' \emph{Physical Review E}, vol.~79, no.~4, p. 041904, 2009.
	
	\bibitem{williams2016dendritic}
	A.~H. Williams, C.~O'donnell, T.~J. Sejnowski, and T.~O'leary, ``Dendritic
	trafficking faces physiologically critical speed-precision tradeoffs,''
	\emph{Elife}, vol.~5, p. e20556, 2016.
	
	\bibitem{kapitein2011way}
	L.~C. Kapitein and C.~C. Hoogenraad, ``Which way to go? cytoskeletal
	organization and polarized transport in neurons,'' \emph{Molecular and
		Cellular Neuroscience}, vol.~46, no.~1, pp. 9--20, 2011.
	
	\bibitem{nirschl2017impact}
	J.~J. Nirschl, A.~E. Ghiretti, and E.~L. Holzbaur, ``The impact of cytoskeletal
	organization on the local regulation of neuronal transport,'' \emph{Nature
		Reviews Neuroscience}, vol.~18, no.~10, p. 585, 2017.
	
	\bibitem{zarai2017deterministic}
	Y.~Zarai, M.~Margaliot, and T.~Tuller, ``A deterministic mathematical model for
	bidirectional excluded flow with langmuir kinetics,'' \emph{PloS one},
	vol.~12, no.~8, p. e0182178, 2017.
	
	\bibitem{doyle2011mechanisms}
	M.~Doyle and M.~A. Kiebler, ``Mechanisms of dendritic mrna transport and its
	role in synaptic tagging,'' \emph{The EMBO journal}, vol.~30, no.~17, pp.
	3540--3552, 2011.
	
	\bibitem{earnshaw2006biophysical}
	B.~A. Earnshaw and P.~C. Bressloff, ``Biophysical model of ampa receptor
	trafficking and its regulation during long-term potentiation/long-term
	depression,'' \emph{Journal of Neuroscience}, vol.~26, no.~47, pp.
	12\,362--12\,373, 2006.
	
	\bibitem{Forni2019}
	F.~{Forni} and R.~{Sepulchre}, ``Differential dissipativity theory for
	dominance analysis,'' \emph{IEEE Transactions on Automatic Control}, vol.~64,
	no.~6, pp. 2340--2351, June 2019.
	
	\bibitem{Miranda-Villatoro2019}
	F.~A. Miranda-Villatoro, F.~Forni, and R.~J. Sepulchre, ``Analysis of lur’e
	dominant systems in the frequency domain,'' \emph{Automatica}, vol.~98, pp.
	76--85, 2018.
	
	\bibitem{Desoer1975}
	C.~Desoer and M.~Vidyasagar, \emph{Feedback Systems: Input-Output Properties},
	ser. Classics in Applied Mathematics.\hskip 1em plus 0.5em minus 0.4em\relax
	Society for Industrial and Applied Mathematics, 1975, vol.~55.
	
	\bibitem{Zhou1995}
	K.~Zhou, J.~Doyle, and K.~Glover, \emph{Robust and optimal control}.\hskip 1em
	plus 0.5em minus 0.4em\relax Prentice Hall, 1995.
	
	\bibitem{VanDerSchaft1999}
	A.~van~der Schaft, \emph{$L_2$-Gain and Passivity in Nonlinear Control},
	2nd~ed.\hskip 1em plus 0.5em minus 0.4em\relax Secaucus, N.J., USA:
	Springer-Verlag New York, Inc., 1999.
	
	\bibitem{reuveni2011genome}
	S.~Reuveni, I.~Meilijson, M.~Kupiec, E.~Ruppin, and T.~Tuller, ``Genome-scale
	analysis of translation elongation with a ribosome flow model,'' \emph{PLoS
		computational biology}, vol.~7, no.~9, p. e1002127, 2011.
	
	\bibitem{spohn2012large}
	H.~Spohn, \emph{Large scale dynamics of interacting particles}.\hskip 1em plus
	0.5em minus 0.4em\relax Springer Science \& Business Media, 2012.
	
	\bibitem{o2014cell}
	T.~O'Leary, A.~H. Williams, A.~Franci, and E.~Marder, ``Cell types, network
	homeostasis, and pathological compensation from a biologically plausible ion
	channel expression model,'' \emph{Neuron}, vol.~82, no.~4, pp. 809--821,
	2014.
	
	\bibitem{o2013correlations}
	T.~O'{L}eary, A.~H. Williams, J.~S. Caplan, and E.~Marder, ``Correlations in
	ion channel expression emerge from homeostatic tuning rules,''
	\emph{Proceedings of the National Academy of Sciences}, vol. 110, no.~28, pp.
	E2645--E2654, 2013.
	
	\bibitem{lmibook}
	S.~Boyd, L.~El~Ghaoui, E.~Feron, and V.~Balakrishnan, \emph{Linear Matrix
		Inequalities in System and Control Theory}.\hskip 1em plus 0.5em minus
	0.4em\relax SIAM, 1994.
	
	\bibitem{padoan2019dominance}
	A.~Padoan, F.~Forni, and R.~Sepulchre, ``The h infinity,p norm as the
	differential gain of a p-dominant system,'' \emph{Accepted to 58th IEEE
		Annual Conference on Decision and Control (CDC)}, 2019.
	
	\bibitem{grant2014cvx}
	M.~Grant and S.~Boyd, ``Cvx: Matlab software for disciplined convex
	programming, version 2.1,'' 2014.
	
\end{thebibliography}

\linespread{0.97}

\end{document}